\documentstyle[titlepage]{article}
\newcommand{\R}{{\rm I\kern-2pt R}}

\begin{document}
\begin{center}
{\bf Comments on `` Quantum Control by Decompositions of $SU(2)$"\\}
Viswanath Ramakrishna\\
Department of Mathematical Sciences and Center for Signals, Systems
and Telecommunications\\
University of Texas at Dallas\\   
Richardson, TX 75083 USA\\
email: vish@utdallas.edu\\
Supported in part by NSF - DMS-0072415 
\end{center}

\noindent{ \bf ABSTRACT:}
The purpose of this note is to i) point out some typographical errors
in \cite{gnvn99an7103}, and ii) to observe that the main result of
\cite{gnvn99an7103} is valid without the restriction that the
matrices $A$ and $B$ in \cite{gnvn99an7103} be orthogonal.

\section{Some Typos and Improvements}

In \cite{gnvn99an7103} a constructive procedure for factorizing any
$SU(2)$ matrix $S$ in the form: 
\begin{equation}
\label{basicdecomp}
S = \Pi_{k=1}^{Q}e^{a_{k}A + b_{k}B}
\end{equation}
with: i) {\bf O1}: $a_{k} > 0$; and ii) {\bf O2}: $\mid b_{k}\mid \leq C$
for an {\em a priori} prescribed bound $C$, was given.  
The only restrictions on the matrices $A$ and $B$ in the published version
of \cite{gnvn99an7103} was that they be i) linearly independent;
and ii) expressible as $A = aX + bY$ and $B = cX + dY$ for any two orthogonal
matrices $X$ and $Y$ in $su(2)$ (the set of all $2\times 2$ skew-Hermitian
matrices with zero trace). Here orthogonality is with respect to the
inner product $ <A, B> = {\mbox Trace} \ (AB^{*})$ (equivalently,
the vectors in $R^{3}$ corresponding to $A$ and $B$ are orthogonal
with respect to the usual inner product on $R^{3}$). {\it There was no
requirement that the vectors $(a, b)$ and $(c, d)$ in $R^{2}$ should
be orthogonal.} 

The first purpose of this note is to observe that the only requirement that
$A$ and $B$ need satisfy is linear independence. Indeed, given two linearly
independent such $A$ and $B$ one can always find (constructively) a unitary
transformation $V\in U(2)$ such that $VAV^{*}$ and $VBV^{*}$ are expressible
as possibly non-orthogonal linear
combinations of $i\sigma_{x}$ and $i\sigma_{y}$. We leave the
technical details to interested readers 
- there are several choices for such $V$'s.
It is useful, in this regard, to note that for any $W\in SU(2)$, the 
transformation $R_{V}: su(2) \rightarrow su(2), R_{V}(A) = VAV^{*}$
is in $SO(3)$ when
$su(2)$ is identified with $R^{3}$.
It then remains to
find  an $SO(3)$ matrix which simultaneously nulls out the $i\sigma_{z}$
component of $A$ and $B$. This can be easily found, for instance,
by modifying the
construction behind Givens rotations. Now any one of the two $SU(2)$
matrices corresponding to this $SO(3)$ matrix will achieve the 
desired purpose. 
Further as $V$ can be found completely constructively,
this coordinate change is {\it known}. So indeed, one can proceed by 
assuming that $A = aX + bY$ and $B = cX + dY$ for any two orthogonal
matrices $X$ and $Y$ in $su(2)$ and still obtain a fully explicit technique 
for preparing any desired unitary transformation in $SU(2)$ via controls
whose pulse area (or power, depending on the interpretation of the $b_{k}$)
is bounded {\it a priori}. The existence of such matrices, $V$, was missed
due to an oversight in \cite{gnvn99an7103} and thus the paper concluded
(incorrectly) that the general case of arbitrary linearly independent $A, B$
perhaps needed more work.
Indeed, all of the above ingredients
were already present in \cite{gnvn99an7103}
and the related paper, \cite{pranew}.
The paper, \cite{thug}, also provides an explicit $V$ which can be used
to null out the $i\sigma_{x}$ component, but without relating it to
$R_{V}$ or Givens rotations.
 
One additional ``improvement" that is immediate in \cite{gnvn99an7103} is that
the values of the number of factors, $Q$, given in Table I on Page 7 can
obviously be lowered, in many instances, by concatenating exponentials
of matrices which are 
constant multiples of each other.
     
The second purpose of this note is to rectify certain typographical errors
which seem to have creeped in during the typesetting process. The main error
is that $\sigma_{z}$ seems to have been replaced by $\sigma_{x}$ at several
points in the published version. These junctures are as follows:

\begin{enumerate} 
\item Everywhere on Page 4, except the headings for subsections
1 and 2 and the headings for Algorithms I and II on Page 4,
$\sigma_{x}$ should read as $\sigma_{z}$.    

\item On Page 5, everywhere in Algorithm II (continued from Page 4) 
$\sigma_{x}$ should be replaced by $\sigma_{z}$.

\item On Page 5, in the heading for Algorithm III,
$\sigma_{x}$ should be replaced by $\sigma_{z}$.

\item In addition,  everywhere on Page 3, $\sigma_{a}$ should  
be replaced by  $\sigma_{z}$. 
\end{enumerate}

\end{document}